\documentstyle[sprocl]{article}

\def\Journal#1#2#3#4{{#1} {\bf #2}, #3 (#4)}

\def\NPB{{\em Nucl. Phys.} B}
\def\PLB{{\em Phys. Lett.}  B}
\def\PRL{\em Phys. Rev. Lett.}
\def\PRD{{\em Phys. Rev.} D}

\def\be{\begin{equation}}
\def\ee{\end{equation}}
\def\bea{\begin{eqnarray}}
\def\eea{\end{eqnarray}}
\newcommand{\ba}{\begin{eqnarray}}
\newcommand{\ea}{\end{eqnarray}}
\newcommand{\Tr}{\mathrm{Tr}}
\def\reff#1{(\ref{#1})}

\def\ltapprox{\raisebox{0.45ex}{$\,\,<$}\raisebox{-0.7ex}{$\!\!\!\!\!\sim\,\,$}}

\def\vsig{\vec{\mbox{$\sigma$}}}
\def\vca{\vec{\mbox{$\cal A$}}}

\begin{document}

\title{THE LATTICE GLUON PROPAGATOR\\ INTO THE NEXT MILLENNIUM}

\author{ATTILIO CUCCHIERI}

\address{Fakult\"at f\"ur Physik, Universit\"at Bielefeld,
33615 Bielefeld, GERMANY\\E-mail: attilio@physik.uni-bielefeld.de}


\maketitle\abstracts{
We evaluate numerically the momentum-space gluon propagator
in the lattice Landau gauge, for three- and four-dimensional
pure $SU(2)$ lattice gauge theory. Although there are large
finite-size effects, we always observe, in the limit of
large lattice volumes, a gluon propagator decreasing in
the infrared limit. This result can be interpreted in a
straightforward way, by considering the proximity of the
so-called first Gribov horizon in the infrared directions.
We also consider the problem of discretization errors
introduced by the lattice regularization, and their effect
on the ultraviolet behavior of the gluon propagator.
}

%%%%%%%%%%%%%%%%%%%%%%%%%%%%%%%%%%%%%%%%%%%%%%%%%%%%%%%%%%%%%%%%%%%%%

\section{Introduction}
\label{sec:intro}

The gluon propagator is not an observable, since it is a
gauge-dependent quantity. Nevertheless, the study of its infrared
behavior provides us with a powerful tool for increasing our
understanding of confinement in non-Abelian gauge 
theories.~\cite{proceeds}
In fact, the infrared
behavior of the gluon propagator can be directly related
to the behavior of the Wilson loop at large separations and
to the existence of an area law.~\cite{West}

The theoretical predictions for the
behavior of the Landau-gauge gluon propagator in the
infrared limit -- obtained
mostly by solving the gluon Dyson-Schwinger equation {\em 
approximately} -- range from a $p^{- 4}$ singularity~\cite{BPenn}
to a vanishing propagator.~\cite{smekal}
A true nonperturbative investigation of these predictions is
possible by performing Monte Carlo simulations of QCD on the
lattice.
 
%%%%%%%%%%%%%%%%%%%%%%%%%%%%%%%%%%%%%%%%%%%%%%%%%%%%%%%%%%%%%%%%%%%%%

\section{The lattice setup}
\label{sec:setup}
 
Let us consider a standard Wilson action for $SU(2)$ lattice gauge
theory in $d$ dimensions. The lattice gluon propagator (in momentum
space) can be written as
\ba
D(0)& \equiv & \frac{1}{d} \sum_{\mu=1}^{d} D_{\mu}(0)
\label{eq:D0def2} \\
D(k) & \equiv & \frac{1}{d-1} \sum_{\mu=1}^{d} D_{\mu}(k)
\;\mbox{,}\;\;\;\;\;\;
\label{eq:Dkdef2}
\ea
where
\be
D_{\mu}(k) \,\equiv\ \frac{\Tr}{6\,V}\,\langle\,
{\widetilde A}_{\mu}(k)\,{\widetilde A}_{\mu}(-k)
\, \rangle \;\mbox{.}\;\;\;\;\;\;
\label{eq:Dkmudef2}
\ee
Here $ V $ is the lattice volume,
\be
{\widetilde A}_{\mu}(k)\,\equiv\,\sum_{x} \, A_{\mu}(x) \,
\exp{\left[ 2 \pi i \left(k \cdot x + k_{\mu}/2\right)\right]}
\ee
and the lattice gluon field $A_{\mu}(x)$ is given by
\be
A_{\mu}(x) \,\equiv\, \frac{1}{2\,i}\,
    \left[\,U_{\mu}(x)\,-\,U_{\mu}^{\dagger}(x)\,\right]
\;\mbox{.}
\label{eq:Amux}
\ee

In order to fix the lattice Landau gauge we can look for a local
minimum of the functional
\be
{\cal E}_{U}[ g ]\,\equiv\,1\,-\,\frac{\Tr}{2\,d\,V}
       \sum_{\mu = 1}^{d}\,\sum_{x}\,
\left[\,g(x)\,U_{\mu}(x)\,g^{\dagger}(x + e_{\mu})\,\right]
\;\mbox{.}
\label{eq:Etomin}
\ee
In fact, if the configuration $\{ U_{\mu}\left(x\right) \}$
is a {\em stationary point} of the functional ${\cal E}_{U}[ g ]$ then
the lattice divergence of $ A_{\mu}(x) $ is zero, i.e.,
\be
\left(\nabla\cdot A \right)(x) \equiv
  \sum_{\mu = 1}^{d} \, \left[ A_{\mu}(x) -
                  A_{\mu}(x - e_{\mu}) \right]
               \, = \, 0 
\;\mbox{.}
\label{eq:diverg0}
\ee
This is the lattice formulation of the usual (continuum) Landau gauge-fixing
condition. Moreover, requiring this stationary point to be a {\em minimum}
of the functional ${\cal E}_{U}[ g ]$ implies that the transverse
gauge-fixed configurations belong to the region $\Omega$ delimited by
the so-called first Gribov horizon, defined as the set of configurations
for which the smallest non-trivial eigenvalue
of the Faddeev-Popov operator is zero.

Thus, when the lattice Landau gauge is imposed, the 
physical configuration space is restricted to the region $\Omega$
and one can prove~\cite{DZ}
a {\em rigorous} inequality for the Fourier components
of the gluon field $ A_{\mu}(x) $.
From this inequality it
follows that the region $\Omega$ is bounded by a certain
ellipsoid $\Theta$. This bound implies the proximity of the first Gribov horizon
in infrared directions and the consequent suppression of the low-momentum
components of the gauge field.
This bound also causes a strong suppression of the gluon propagator
in the infrared limit. In fact, Zwanziger proved~\cite{Z1}
that, in four dimensions and in the infinite-volume limit, the gluon propagator
is less singular than $p^{-2}$ in the infrared limit and that, very likely, it
{\em does} vanish in this limit.
A similar result holds in three dimensions: one obtains
that, in the infinite-volume limit,
the gluon propagator must be less singular than $p^{-1}$ as $p \to 0$ and that,
very likely, it vanishes in the infrared limit.
We remark that these predictions
for the gluon propagator are {\em $\beta$-independent}: in fact,
they are derived only from the positiveness of the Faddeev-Popov
operator when the lattice Landau gauge is imposed.

%%%%%%%%%%%%%%%%%%%%%%%%%%%%%%%%%%%%%%%%%%%%%%%%%%%%%%%%%%%%%%%%%%%%%

\section{The infrared behavior of the gluon propagator}
\label{sec:infra}

We have studied the momentum-space gluon propagator $ D(k) $
in four~\cite{Attilio4d}
and in three~\cite{Attilio3d}
dimensions.
In both cases we have found that, if the lattice volume $ V $ is large enough,
the gluon propagator is decreasing as the magnitude of the lattice momentum
$ p(k)
\equiv
2\,\left[\,\sum_{\mu = 1}^{d} \sin^{2}{\left( \pi \,k_{\mu} \right)}\,
\right]^{1/2} $
decreases, provided that $p(k)$ is smaller than a value $p_{dec}$.
Also, the lattice volume at which this behavior for the gluon
propagator starts to be observed increases with the coupling
$\beta$, i.e., finite-size effects are very large in the
small-momenta sector. This makes practically unfeasible, with present 
computational resources, the numerical study of the infrared behavior
of the gluon propagator in four dimensions and at large values of
$\beta $, and explains why a decreasing gluon propagator has
been observed only in the strong-coupling regime for the four-dimensional 
case.~\cite{Attilio4d,Naka}

In the three-dimensional case we get good scaling for values
of $ \beta $ greater than $ 3.4 $, especially in the
region of large momenta, where finite-size effects are
negligible.
We also see that the gluon propagator is decreasing for momenta
$p \ltapprox p_{dec}$, and that the value of $p_{dec}$ (in physical units)
is practically $\beta$-independent. From our data we obtain
$p_{dec} \approx 350 \, \mbox{MeV} $.
For the same set of data we also observe that the gluon
propagator is less singular than $p^{-1}$ in the infrared limit,
in agreement with Zwanziger's prediction. We notice that
the {\em turnover} momentum $p_{to} \approx 700 \, \mbox{MeV} $ is in good
agreement with the result obtained recently in four dimensions for the
$SU(3)$ group.~\cite{par}
Finally, the value
$D(0)$ of the gluon propagator at zero momentum decreases monotonically as
the lattice volume increases (see for example the case $\beta = 5.0$ in the
three-dimensional case~\cite{Attilio3d}).
This suggests a finite value for $D(0)$ in the
infinite-volume limit, but it is not clear whether this value would be zero or a
strictly positive constant. Therefore, the possibility of a zero value for
$D(0)$ in the infinite-volume limit is not ruled out.

%%%%%%%%%%%%%%%%%%%%%%%%%%%%%%%%%%%%%%%%%%%%%%%%%%%%%%%%%%%%%%%%%%%%%

\subsection{Conclusions}

The prediction~\cite{smekal,Z1}
of a gluon propagator decreasing for momenta $p(k) \ltapprox p_{dec}$
is clearly verified numerically for several values of the coupling
$\beta$. In the three-dimensional case these values range from the
strong-coupling regime to the scaling region.
Also, our data in the strong-coupling regime for
the three-dimensional case are in qualitative agreement with the 
results obtained in four dimensions.~\cite{Attilio4d}
This strongly suggests to us that a similar analogy will hold
--- in the limit of large lattice volumes --- for couplings $\beta$
in the scaling region, leading to an infrared-suppressed gluon
propagator also in the four-dimensional case.

%%%%%%%%%%%%%%%%%%%%%%%%%%%%%%%%%%%%%%%%%%%%%%%%%%%%%%%%%%%%%%%%%%%%%

\section{Discretization effects}

The definition of the lattice gluon field given in
Eq.\ \reff{eq:Amux} is only one of the possible lattice discretizations
of $ A_{\mu}(x) $, i.e., we can consider several possible
definitions, leading to discretization errors of different orders.
For example, we can write~\cite{CMHT}
\vspace{-2mm}
\bea
A_{\mu}^{(1)}(x) &\equiv&
 \frac{U_{\mu}(x)\,-\,U_{\mu}^{\dagger}(x)}{2\,i} \\[1mm]
 A_{\mu}^{(2)}(x) &\equiv&
 \frac{[U_{\mu}(x)]^2\,-\,[U_{\mu}^{\dagger}(x)]^2}{4\,i} \\[1mm]
 A_{\mu}^{(3)}(x) &\equiv&
 \frac{[U_{\mu}(x)]^4\,-\,[U_{\mu}^{\dagger}(x)]^4}{8\,i}\;\;
\mbox{.}
\eea
If we set
\be
U_{\mu}(x)\equiv \exp[i a g_0 \,\vsig \cdot \vca(x)]
\;\mbox{,}
\ee
we obtain that $A_{\mu}^{(1)}(x)$, $A_{\mu}^{(2)}(x)$ and
$A_{\mu}^{(3)}(x)$
are equal to $\, a g_0 \,\vsig \cdot \vca(x) $ plus
terms of order $a^3\,g_0^3$. We can also consider
\bea
A_{\mu}^{(4)}(x) & \equiv & \left[\,
  4\,A_{\mu}^{(1)}(x) \,-\,
   A_{\mu}^{(2)}(x) \,\right]\,/\,3 \\
A_{\mu}^{(5)}(x) & \equiv & \left[\,
  16\,A_{\mu}^{(1)}(x) \,-\,
   A_{\mu}^{(3)}(x) \,\right]\,/\,15 \\
A_{\mu}^{(6)}(x) & \equiv & \left[\,
  4\,A_{\mu}^{(2)}(x) \,-\,
   A_{\mu}^{(3)}(x) \,\right]\,/\,3 \;
\eea
    and
\be
    A_{\mu}^{(7)}(x) \;\equiv\;\left[\,
    64\,A_{\mu}^{(1)}(x) \,-\,
   20\,A_{\mu}^{(2)}(x) \,+\,
    A_{\mu}^{(3)}(x)\,\right]\,/\,45 \;.
\ee
It is easy to check that $A_{\mu}^{(4)}(x)$, $A_{\mu}^{(5)}(x)$ and
$A_{\mu}^{(6)}(x)$
are equal to $\, a g_0 \,\vsig \cdot \vca(x) $ plus
terms of order $a^5\,g_0^5$, and that
$A_{\mu}^{(7)}(x) = a g_0 \,\vsig \cdot
\vca(x)$ plus terms of order $a^7\,g_0^7$.

The definitions $A^{(1)}$ and $A^{(2)}$ were recently
considered by Giusti {\em et al.}~\cite{Giusti}
They find that the corresponding gluon propagators are equal modulo a
constant factor. We have performed a similar study~\cite{CMHT}
at several values of $\beta$ and for lattice volumes $V=8^4$
and $12^4$,
evaluating $D^{(i)}(k)$ using the different definitions
of the gluon field $A_{\mu}^{(i)}$ given above.
We obtain, in all cases, that the
seven propagators $D^{(i)}(k)$ are equal modulo a
constant factor.
We notice that this proportionality constant
between different discretizations of the
gluon propagator may be explained as a tadpole
renormalization. In fact, let us consider the
tadpole-improved link ${\widetilde U}_{\mu}(x)\equiv U_{\mu}(x)/u_0$,
where $u_0$ is the mean link in Landau gauge.
Then $D^{(1)}(k)$ gets multiplied by a factor
$u_0^{-2}$, and $D^{(2)}(k)$ by $u_0^{-4}$.
For example, for $V=12^4$ and $\beta=2.2$ we have $u_0^{2} = 0.68428(8)$.
In this case, using tadpole-improved operators,
the discrepancy $D^{(1)}(k)/D^{(2)}(k)$
is reduced from $1.833(4)$ to $1.254(3)$. Similarly, at $\beta=2.3$ [and $2.7$],
we obtain $u_0^{2} = 0.7222(1)$ [respectively $u_0^{2} = 0.8077(3)$] and
the discrepancy $D^{(1)}(k)/D^{(2)}(k)$
is reduced from $1.701(4)$ [respectively $1.425(3)]$
to $1.228(3)$ [respectively $1.151(2)$]. As expected, the
discrepancy between different discretizations decreases as
$\beta$ increases.

\begin{table}[t]
\caption{The parameter $Z$ and the $\chi^{2}/$d.o.f.\ for the best fit to
the ultraviolet behavior $ Z g_{0}^{2} / \left[ 4\, p^{2}(k) \right] $.
We consider different lattice
discretizations of the gluon propagator, and indicate with
$D^{(i)}_{TI}(k)$ the $D^{(i)}(k)$ propagator
evaluated using tadpole-improved links ${\widetilde U}_{\mu}(x)$.
\label{tab:UV}}
\begin{center}
\footnotesize
\begin{tabular}{|c||c|c||c|c|}
\hline
& \multicolumn{2}{|c||}{\raisebox{0pt}[9pt][5pt]{$\beta = 2.7$}} &
    \multicolumn{2}{|c|}{\raisebox{0pt}[9pt][5pt]{$\beta = 10$}} \\
\cline{2-5}
\begin{minipage}{0.7in}
\protect\vspace{-0.4cm}
\begin{center}
Type of propagator
\end{center}\end{minipage}
& \raisebox{0pt}[11pt][7pt]{$Z$}
& \raisebox{0pt}[11pt][7pt]{$\chi^{2}/$d.o.f} & \raisebox{0pt}[11pt][7pt]{$Z$} &
\raisebox{0pt}[11pt][7pt]{$\chi^{2}/$d.o.f} \\
\hline
\raisebox{0pt}[11pt][6pt]{$D^{(1)}(k)$} & 1.2965 & 1.36 & 1.0261 & 0.121 \\[3pt]
$D^{(1)}_{TI}(k)$ & 1.6052 & 1.36 & 1.0904 & 0.121 \\[3pt]
$D^{(2)}(k)$ & 0.9141 & 1.25 & 0.9295 & 0.133 \\[3pt]
$D^{(2)}_{TI}(k)$ & 1.4012 & 1.25 & 1.0496 & 0.133 \\[3pt]
\hline
\end{tabular}
\end{center}
\end{table}

It is also interesting to compare the data obtained for the gluon
propagator at large momenta with the ultraviolet behavior
$ g_{0}^{2} / \left[ 4\, p^{2}(k) \right]$,
predicted by perturbation theory at zeroth order.
To this end, we consider different discretizations of
the gluon propagator and we fit the data corresponding to $p^{2}(k) \geq 3$
using the function $ Z g_{0}^{2} / \left[ 4\, p^{2}(k) \right] $.
In Tab.\ \ref{tab:UV} we report the results obtained
for the lattice volume $V=12^4$ at $\beta = 2.7$ and $\beta = 10$.
Again, the discrepancy between $D^{(1)}(k)$ and $D^{(2)}(k)$
is reduced when tadpole-improved operators $D^{(i)}_{TI}(k)$
are used.
Also, as expected, the value of $Z$
gets closer to $1$ as $\beta$ increases.

Finally, let us notice that, in finite-temperature QCD,
the long-distance behavior of the gluon propagator is directly
related to the electric and magnetic screening lengths, and
that these screening masses are invariant under rescaling of the
propagators by a constant factor.
In particular, $D_{1}^{(i)}(k) \,+\, D_{2}^{(i)}(k)$
[respectively $D_{4}^{(i)}(k)$] is related to the gluon propagator
used by Karsch {\em et al.}~\cite{Karsch}
for the evaluation of the magnetic [respectively electric]
screening mass. It has been checked~\cite{CKHT}
that, also at finite temperature, different discretizations
for the gluon propagator are equal modulo a constant value, i.e.,
the screening masses are independent of the discretization $D^{(i)}(k)$.

%%%%%%%%%%%%%%%%%%%%%%%%%%%%%%%%%%%%%%%%%%%%%%%%%%%%%%%%%%%%%%%%%%%%%

\section*{Acknowledgments}
This work was partially supported by the TMR network Finite Temperature
Phase Transitions in Particle Physics, EU contract no.: ERBFMRX-CT97-0122.

%%%%%%%%%%%%%%%%%%%%%%%%%%%%%%%%%%%%%%%%%%%%%%%%%%%%%%%%%%%%%%%%%%%%%

\end{document}